\documentclass[a4paper,12pt]{article}
\usepackage[brazil]{babel}
\usepackage[latin1]{inputenc}
\usepackage{amsthm,amsfonts}
\usepackage[normalem]{ulem}
\usepackage{epsfig}
\usepackage{graphicx}
\begin{document}
\begin{center}\LARGE\textbf{Factorization of large numbers and the suggestion of an algorithm}\end{center}
\begin{center}\large\textbf{Fabiano Sutter de Oliveira}\footnote{CNPq Fellow of the PIBIC at CBPF}\end{center}
\begin{center}Centro Brasileiro de Pesquisas Físicas - CBPF\footnote{This work has been presented in the Poster Session of the XII Jornada Científica of the PIBIC-CBPF, October 2005.} - LAFEX\\Rua Dr. Xavier Sigaud, 150 - Urca - Rio de Janeiro - RJ\end{center} 

\begin{center}\texttt{fabianosutter@cbpf.br}\end{center}
\begin{quote}\textit{\bf Abstract}: \textit{In this paper, we intend to present a new algorithm to factorize large numbers. According to the algorithm proposed here, we prove that there is a common factor between p and q. With this procedure, the time of factorization considerably decreases. The algorithm is based on a graphic representation and, when the corresponding graph is drawn, coordinate pairs will ori-\\ginate two straight lines that intercept one another. These coordinate pairs are formed by prime numbers in the x-axis, and  factors in the y-axis, including the factor in common.}\end{quote}
\begin{quote}\textit{\bf Keywords}: \textit {Cryptography. Factorization. Algorithm. RSA.}\end{quote}

\section{Introduction}
  \ \ \ \ The main scope of Cryptography is to provide security and protection for the transmission of data, with the care of ensuring secrecy and integrity of the information up to its final address. The difficulty of factorization is fundamental for the security of the messages that travel every day through Internet, and the study of algorithms deviced to factorize these numbers is a matter that grows in importance as time goes by.\\

One of the cryptosystems most currently adopted is the so-called RSA, developed in 1978, at MIT, by R. Rivest, A. Shamir and L. Adleman. It supports almost 95\% of all e-commerce. Based on the multiplication of two prime numbers, through some mathematical operations, it ciphers the messa-\\ges, before they are sent to their destination. The inverse process, i.e., the decodification, becomes impossible with the existing technology at our disposal. According to calculations, the time for the factorization of a value of 1024 bits, for example, would take a hundred thousand years. The security of RSA depends on the difficulty of factorizing these numbers. The task is complex even with the presently powerful computers.\\

In 1999, after 7 months of studies, a key of the RSA-512 bits was broken by researchers in Netherlands, with the aid of others six countries and 300 workstations. These facts are very disturbing, but the breakdown of this key, like other keys, still represents a difficult task for a normal user, because one could need powerful computers, working during many years.\\

With the advent of the Quantum Computer, it will be unavoidable to create new forms of Cryptography. Quantum computers are faster than their classical partners. In short time, we will be able to factorize a key with any number of bits in minutes, provided that a quantum computer is used.\\

One of the questions associated to the factorization of large numbers is the time necessary for its factorization. In other words, the time for carrying out this task must be measured and classified in polynomial time or exponential time. The problems of exponential order are practically impossible to be solved, even with the most advanced computers. All the algorithms created up to now, except the quantum ones, fall down into the exponential time category. Perhaps, an appropriate algorithm can developed to accomplish the task of factorization in a polynomial time. Thus, mathematicians would prove that P\footnote{Polynomial-time} and NP\footnote{Not polynomial-time} are equivalent and this would imply to say that the problem to decipher codes could be decided in a polynomial time.\\

\ In this paper, we intend to present a new algorithm to factorize large numbers and we shall prove that there is a factor in common, between \textit{p} and \textit{q}, according to the algorithm presented. There is a limit range for the search, diminishing then the worktime. The method employed is based on a property that states and proves that there is a factor in common between \textit{p} and \textit{q} that is the same for both. When the graph is drawn, the coordinate pairs originate two straight lines that intercept one another. These coordinate pairs are formed by prime numbers, in the x-axis, and factors, in y-axis, including the factor in common.\\
\newpage
This paper is outlined as follows:\\

In Section 2, we present the general scheme of the algorithm we propose here;\\

next, in Section 3, we provide an example to illustrate how the method  works;\\

finally, in Section 4, we draw our General Conclusions and discuss some possibilities of new developments.\\

An appendix follows where the results of a partial factorization of the RSA-640 and the graph (Fig. 1) are shown.\

\section{General scheme}

  \ \ \ \ We intend to present here how the process is initiated. The operations discussed in this process prepare the value of \textit{N} (the number to be factorized) for the algorithm we work out in the sequel.\\

Suppose we have a set U capable to generate the factors of \textit{N}:
\begin{eqnarray}
U=\{p\in \mathbb{N}|p< \sqrt{N}\}
\end{eqnarray}
\begin{eqnarray}
U=\{q\in \mathbb{N}|q< \sqrt{N}\}
\end{eqnarray}
  \ \ \ \ Presently, one knows that both the factors are smaller than the square root of \textit{N}.\\

Whenever $\sqrt{N}$ is calculated, we can say that the search for \textit{p} and \textit{q} is over:
\begin{eqnarray}
p\times q=N                                  
\end{eqnarray}
\begin{eqnarray}
\overline{p+q}=g                           
\end{eqnarray}
\begin{eqnarray}
g^2\approx N
\end{eqnarray}

 \ In the previous equations, one gets approximately the value of \textit{N}, when we calculate the average of the product of two arbitrary values.\\

For the process of factorization, the known relations are the following:
\begin{eqnarray}
N=p\times q
\end{eqnarray}
\begin{eqnarray}
\frac{N}{p}=q
\end{eqnarray}
\begin{eqnarray}
\frac{q}{N}=\frac{1}{p}
\end{eqnarray}
\begin{eqnarray}
\sqrt{N}={\kappa} 
\end{eqnarray}
\
  \ \ \ \ Reformulating these equations, we can write:
\begin{eqnarray}
\frac{\kappa}{p}=\textit{f}
\end{eqnarray}
\begin{eqnarray}
\frac{q}{\kappa}=\textit{f}
\end{eqnarray}

The equations (10) and (11) prove that there exists a common factor \textit{f}, found in both the quotients. Each factor of the product will have its corresponding interval; we shall denote them by \textit{a} and \textit{b}. Generally, the following intervals are found:
\begin{eqnarray}
a=[0.01,1]
\end{eqnarray}
\begin{eqnarray}
b=[1,2]
\end{eqnarray}
\

 The intervals will help in the restriction of the search, in that they fix an initial and a final value for each equation.\\

If a factor exists in common, we conclude that, by using straight lines and their intersection, this common factor may be found in a straight forward way. The exactness increases whenever we come close to this factor.\\

With theses values, a graph may be drawn where in the x-axis we have prime numbers, and in the y-axis we find the factors. In this moment, the straight lines intercept each other.\
\newpage
\section{Examples and applications}

  \ \ \ \ We shall here explain, by means of a general example, how the algorithm actually works.\\

Taking as an example (p = 47809 ; q = 78437):
\begin{eqnarray}
N=3749994533 = 47809\times 78437
\end{eqnarray}

We get\begin{eqnarray}\kappa \cong 61237\end{eqnarray}

To separate the field of search in two groups, we take the product between (15) and the intervals given in (12) and (13). Taking (10) and (11), we find a group of dispersed factors, but the common factor will always exist. For this set, it is 1.2808/1.2809. Both the straight lines converge to a common point, that is, the factor \textit{f}.\\

Some particularitities of \textit{f} must be noticed. Among them, the existent number of digits in common when we get close to the common factor.\\

The factor is always inside an interval, as it can be seen, from (12) and (13). Within this interval, one knows that the beginning and the end of the search are between 0.01 (16) and 2 (17):
\begin{eqnarray}
p=[\kappa\times 0.01,\kappa\times 1]
\end{eqnarray}
\begin{eqnarray}
q=[\kappa\times 1.01,\kappa\times 2]
\end{eqnarray}

Once it is proven that there exists an identical factor among the factors of \textit{N}, a number of suitable relations may be found. We think it is not worthy to quote them here.\\

Our conclusions is that to find the factors, we have two search universes. Each search universe is responsible for a straight line, as it is depicted in the graph of Fig. 1.\\
\newpage
\section{General conclusions}

\ \ \ The algorithm described above has not yet been implemented in any programming language. It has already been tested with Maple, to ascertain that the equations we found above are valid. The results are promising. The proposal of the paper is, besides presenting an algorithm, stimulating the compilation of the full algorithm or part of it and going through new studies that involve other variants of the process. We present, still in this paper, a demonstration that the algorithm would work. For that, we use the factori-\\zation of RSA-640, by now already factored, but, at the time of our study, it was still open. The presentation (Appendix) demonstrates only the beginning of the way the algorithms works, using RSA-640, like an example for factorization. The breaking of the RSA-640 was reported on November 2, 2005, after the content of this factorization was presented at the Poster Session of the XII Jornada Científica do PIBIC-CBPF, on October 28, 2005.\\

\section{Acknowledgments}

  \ \ \ \ The author would like to thank Prof. J.A. Helayel Neto for discussions and a careful reading of the original manuscript. He also expresses his gratitude to CNPq - Brasil for his PIBIC - fellowship.\

\section{References}

  \ \ \ \ [1] Feynman, R. P. (1982), {Simulating physics with computers, International Journal of Theoretical Physics 21, 467-488}\\

[2] Bringsjord, Selmer and Taylor, Joshua (2004), {N=NP, arXiv:cs.0406056}\\

[3] Galindo, A., and Martin- Delgado, M. A. (2001), {Information and Computation: Classical and Quantum Aspects, arXiv: quant-ph/0112105 v1}\\

[4] Lavor, C., Mansur, L. R. U., Portugal, R. (2003), {Shor's Algorithm for Factoring Large Numbers, arXiv: quant-ph/0303175 v1}\\

[5] Takagi, Nilton Hidek (2003), {Fundamentos Matemáticos de Criptografia Quântica - Monografia - Universidade Federal do Mato Grosso}\\

[6] RSA Security (2005-2006), {The RSA Challenge Numbers. Disponível em: http://www.rsasecurity.com} - Acesso: 25 Out. 2005.\\

[7] Wolfram Institute, {RSA-640 Factored - Disponível em: http://mathworld.\\wolfram.com/news/2005-11-08/rsa-640} - Acesso: 09 Nov. 2005.

\newpage
\begin{center}\huge\textbf{Appendix}\end{center}
Factorization of RSA-640 - RSA\\
\\
\uline{31074182404900437213507500358885679300373460228427275457201619488\\
2320644051808150455634682967172}3286782437916272838033415471073108\\
501919548529007337724822783525742386454014691736602477652346609\\
\\\
p = 1634733645809253848443133883865090859841783670033\\
092312181110852389333100104508151212118167511579\\
\\
q = 1900871281664822113126851573935413975471896789968\\ 
51549366663853908802710380210449895719126.1465571\\
\\
\textit{f} = 1.07833289578512094427148576077075973244697825135682087002279\\
6256440687462122306160642947834542600419\\
\\
\textit{f} = 1.07833289578512094427148576077075973244697825135682087002279\\
6256440687462122306160642947834542600688\\
\\
Factorization of RSA-640 - Algorithm - Partial factorization\\
\\
\uline{31074182404900437213507500358885679300373460228427275457201619488\\
2320644051808150455634682967172}0043741354331993532308323732530432\\
562647669101807527524162144028484341037141008649554752230426788\\
\\\
p = 1410229652898288755049364589420981108074639325137\\
685864098536262301424020609610329704498357227684\\
\\
q = 2203483832653576179764632170970282981366623945527\\
634162653962909845975032202516140163278683168257\\
\\
$\kappa\cong 1762787066122860943811705736776226385093299156422107330123170\\
327876780025762012912130622946534606$\\

\newpage

\begin{center}\huge\textbf{Figure 1}\end{center}

\center{\psfig{figure=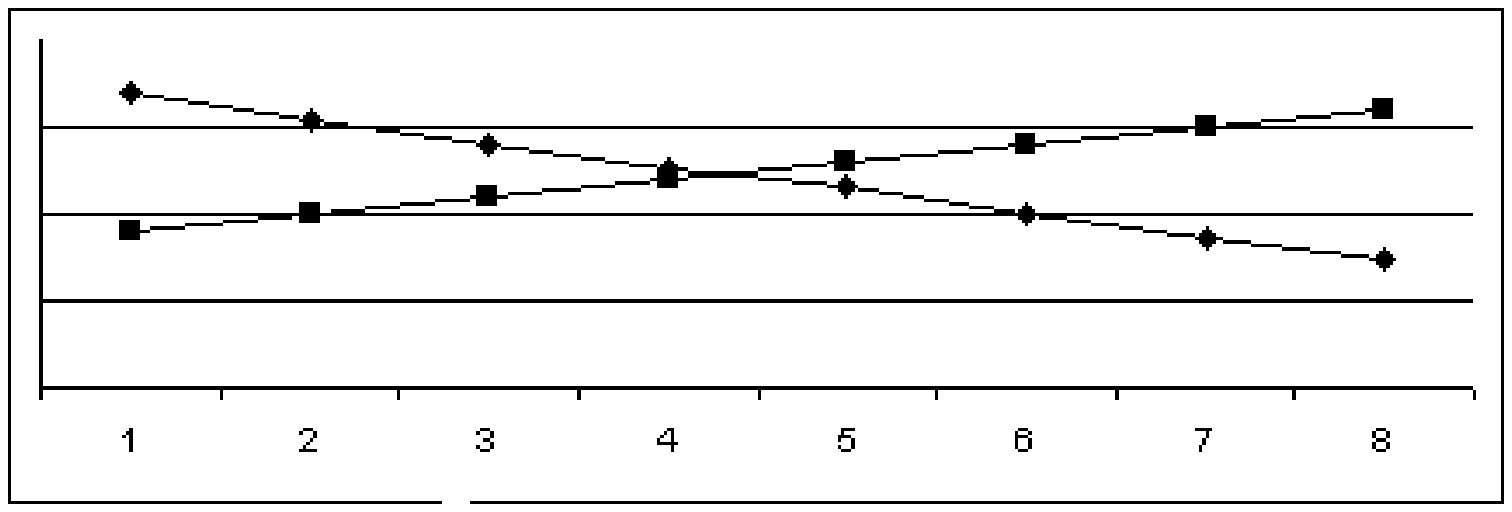,height=13cm,width=17cm}}

\end{document}